# Anti-Sampling-Distortion Compressive Wideband Spectrum Sensing for Cognitive Radio

Yipeng Liu[†], Qun Wan[†]

[†] Department of Electronic Engineering, University of Electronic Science and Technology of China (UESTC), Chengdu 611731, China

{liuyipeng; wanqun}@uestc.edu.cn

*Abstract* — Too high sampling rate is the bottleneck to wideband spectrum sensing for cognitive radio in mobile communication. Compressed sensing (CS) is introduced to transfer the sampling burden. The standard sparse signal recovery of CS does not consider the distortion in the analogue-to-information converter (AIC). To mitigate performance degeneration casued by the mismatch in least square distortionless constraint which doesn't consider the AIC distortion, we define the sparse signal with the sampling distortion as a bounded additive noise, and An anti-sampling-distortion constraint (ASDC) is deduced. Then we combine the $\ell_1$ norm based sparse constraint with the ASDC to get a novel robust sparse signal recovery operator with sampling distortion. Numerical simulations demonstrate that the proposed method outperforms standard sparse wideband spectrum sensing in accuracy, denoising ability, etc.

I. INTRODUCTION

Current static spectrum assignment policy leads to a paradox between the overcrowded spectrum and the pervasiveness of idle frequency bands [1]. To deal with the problem, a fast developing revolutionary technique named cognitive radio (CR) is proposed to achieve more flexible spectrum management policies and technologies for dynamic radio spectrum access [2]-[3]. CR is defined as a paradigm for wireless communication in which either a network or a wireless node rapidly tunes its transmission or reception parameters to communicate efficiently avoiding interference to and from licensed or unlicensed users. This alteration of parameters is based on the dynamical monitoring of the external and internal radio environment, such as radio frequency spectrum, user behavior and network state. It is regarded as the next big thing in wireless communication.

Spectrum sensing is an essential function of CR. It autonomously detects the spectrum usage without harmful interference, and then the CR users can be allowed to utilize the unused primary frequency bands. Current spectrum sensing is usually performed in two steps [4]: the first step, called coarse spectrum sensing, is to detect the power spectrum density (PSD) level of primary bands; and the second step, called feature detection or multi-dimensional sensing [5] [6], is to estimate other signal space allowing accessible for CR, such as direction of arrival (DOA) estimation, spread spectrum code identification, waveform identification, etc.

Coarse spectrum sensing requires fast and accurate power spectrum detection over a wideband and even ultra-wideband (UWB) for many CR missions [7]. One approach utilizes a bank of tunable narrowband bandpass filters to search one frequency band at a time. However, it requires an enormous number of RF components, which leads to high cost and much time-consuming. The other one is a wideband circuit using a single RF chain followed by high-speed DSP to flexibly search over multiple frequency bands concurrently [8]. High sampling rate requirement and the resulting large number of data for processing are the major problems [9].

Compressive sensing (CS) is a very promising technique to deal with these problems [10][11][12][13]. CS is a new signal processing technique that can reconstruct the signal with a much fewer randomized samples than Nyquist sampling with high probability on condition that the signal has a sparse representation. The sampling measurement matrix should be incoherent with the dictionary. Orthogonal matched pursuit (OMP) [14], least-absolute shrinkage and selection operator (LASSO) [15] and Dantzig Selector (DS) [16] are the major ways to recovering the sparse signal.

The standard sparse signal recovery algorithms, such as BP, were used to recover the spectrum from the random samples obtained by analog to information converter (AIC) [10][11] as Fig. 1 shows. In [9], an analog-to-digital converter has been used to transform the analog received signal into a digital signal by sampling at the Nyquist rate. Next, sub-Nyquist random sampling is applied to the sampled vector to compress it into a shorter vector and then the power spectrum is reconstructed by solving the BP problem. In [10][11] the received analog signal can be sampled at the information rate of the signal using an AIC. Here, the CS is embedded in the AIC. The same sparse signal recovery methods can be used to estimate the original power spectrum density (PSD). In [9], wavelet edge detection has been used to detect the spectrum borders in the estimated spectrum and the detection's performance has been evaluated in terms of mean square error through simulations.

However, traditional methods do not consider uncertainties in AIC. In fact, sampling distortion can be resulted from various types of effects, such as aliasing, aperture effect, jitter or deviation from the precise sample timing intervals, noise, slew rate limit error, and other non-linear effects [17]. Sampling distortion would change the compressive sampling signal model and make the present sparse signal recovery methods mismatch the signal. Thus, the performance of the CWSS would be deteriorated.

To prevent performance degradation caused by sampling distortion, this paper proposes a robust compressive wideband spectrum sensing method. Sampling distortion is modeled as a bounded additive error. The estimated signal approximates the sparse signal with a mixed $\ell_1$ and $\ell_2$ norms with respect to the sampling distortion parameter. Then we combine the sparse constraint with the mixed $\ell_1$ and $\ell_2$ norms constraint to get the robust compressive wideband spectrum sensing operator. Numerical simulations demonstrate that the proposed operator has a better reconstructing performance with the sampling distortion than traditional methods.

In the rest of the paper, the CS theory is reviewed briefly in Section II. Section III gives the signal model. Section IV provides a typical representative of the traditional compressive wideband spectrum sensing method; In Section V, the proposed robust wideband spectrum sensing against sampling distortion is given. In Section VI, performance enhancement of the proposed operator is demonstrated by numerical experiments. Finally, Section VII draws conclusion and discusses on further work.

## II. COMPRESSED SENSING REVIEW

Sparsity widely exists in wireless signals [18]. Considering a signal **x** which can be expanded by an orthogonal complete dictionary with the representation given as

$$\mathbf{x}_{N\times 1} = \mathbf{\Psi}_{N\times N}\mathbf{b}_{N\times 1}, \qquad (1)$$

when most elements of the vector **b** are zeros, the signal **x** is sparse. If the number of nonzero elements of **b** is S (S<<N), the signal is said to be *S*-sparse. CS provides an alternative to the well-known Shannon sampling theory. On condition that the signal is sparse, CS performs non-adaptive measurement of the informative part of the signal directly.

In CS, instead of measure the signal directly by Nyquist sampling, a random measurement matrix $\mathbf{\Phi}$ is used to sample the signal. In matrix notation, the obtained random sample vector can be represented as:

$$\mathbf{y}_{M\times 1} = \mathbf{\Phi}_{M\times N}\mathbf{x}_{N\times 1}, \tag{2}$$

The measurement matrix should satisfy the restricted isometry property (RIP) which provides a guarantee on the performance of $\mathbf{\Phi}$ in CS. It can be stated as [12][13][18]:

$$(1-\delta_s)\|\mathbf{y}\|_2^2 \leq \|\mathbf{\Phi}\mathbf{y}\|_2^2 \leq (1+\delta_s)\|\mathbf{y}\|_2^2, \tag{3}$$

for all the *S*-sparse $\mathbf{y}$. In (3), $\|\mathbf{y}\|_2 = \left(\sum_i |y_i|^2\right)^{1/2}$ is the $\ell_2$ norm of the vector $\mathbf{y}$ = [y$_1$, y$_2$, ... , y$_M$]$^T$. The restricted isometry constant $\delta_s \in (0,1)$ is defined as the smallest constant for which this property holds for all S-sparse vectors $\mathbf{y}$.

There are three kinds of frequently used measurement matrices:

1) Non-Uniform Sub-sampling (NUS) or Random Sub-sampling matrices which are generated by choosing *M* separate rows uniformly at random from the unit matrix $\mathbf{I}_N$;

2) Matrices formed by sampling the i.i.d. entries $(\mathbf{\Phi})_{ij}$ from a white Gaussian distribution;

3) Matrices formed by sampling the i.i.d. entries $(\mathbf{\Phi})_{ij}$ from a symmetric Bernoulli distribution and the elements are $\pm 1/\sqrt{N}$ with probability 1/2 each.

The hardware performing random samplings of type 1 and 3 can refer to [10][11]. When the RIP holds, a series of recovering algorithm can reconstruct the sparse signal. One is greedy algorithm, such as matched pursuit (MP), OMP; another group is convex program, such as basis pursuit (BP), LASSO and Dantzig Selector (DS). In these algorithms, DS has almost the same performance as LASSO. Convex program algorithm has a more reconstruction accuracy while greedy algorithm has less computational complexity.

## III. Signal Model

According to the measured data [1][9], we may suppose that the receiver provides a wideband monitoring frequency range and the received signal, $y(t)$, $0 \leqslant t \leqslant T$, just occupies a small parts of non-consecutive frequency bands with different PSD levels. It can be represented as

$$y(t) = \sum_{i=1}^{N} c_i \psi_i(t), \qquad (4)$$

where $\Psi_i(t)$ is the Fourier basis functions and $c_i$ is the weighting factors indicating the PSD. In the weighting vector $\mathbf{c} = [c_1, c_2, \ldots, c_N]^T$, most of the elements are zeros or trivial.

### A. Traditional sparse spectrum model

In traditional Nyquist sampling, the time window for sensing is $t \in [0, NT_0]$. $N$ samples are needed to recover the PSD $r$ without aliasing, where $T_0$ is the Nyquist sampling duration. A digital receiver converts the continuous signal $y(t)$ to a discrete complex sequence $y$ of length M. For illustration convenience, we detail the algorithm in discrete setting as it did in [3], [19]-[24]:

$$\mathbf{y} = \mathbf{\Psi} \mathbf{y}_t, \mathbf{y}_t \in \mathbf{R}^N, \qquad (5)$$

where $\mathbf{y}_t$ represents an $N \times 1$ vector with elements $y_t[n] = x(t)$, $t = nT_0$, $n = 1, \ldots, N$, and $\mathbf{S}$ is an $M \times N$ projection matrix. For example, when $\mathbf{\Psi} = \mathbf{F}_N$, model (5) amounts to frequency domain sampling, where $\mathbf{F}_N$ is the $N$-point unitary discrete Fourier transform (DFT) matrix. Given the sample set $y$ when $M < N$, compressive spectrum sensing can reconstruct the spectrum with reduced amount of sampling data.

Processing $y(t)$ in such a broad band needs high sampling rate which has high cost. Besides, too many sampling measurements inevitably increase computational burden for digital signal processors. Nevertheless, spectrum sensing requires a fast and accurate algorithm. Considering $y_t$ has a sparse representation in frequency domain, we use an $M \times N$ random projection matrix $\boldsymbol{S}_c$ to sample signals as the type 1 random measurement in section 2, i.e. $\boldsymbol{y} = \boldsymbol{S}_c \boldsymbol{y}_t$, where $M < N$.

In practice, the analog baseband signal $y(t)$ is sampled using an analog-to-information converter (AIC). The AIC can be conceptually modeled as an ADC operating at Nyquist rate, followed by CS operation. Then $\boldsymbol{y}$ is obtained directly from continuous time signal $y(t)$ by AIC [10][11]. Here we incorporate the AIC to the wideband spectrum sensing architecture as Fig. 1.

B. *Practical sparse spectrum model with sampling distortion*

The AIC is a non-ideal device with various physical limitations. Various types of effects can lead to sampling distortions, including:

1) Aliasing. The ideal sampling theorem assumes that the signal is bandlimited. However, in practice, bandlimitated signal cannot be guaranteed for the time-limited signal. Since the sampled continuous signals are almost time-limited (e.g., at most spanning the lifetime of the sampling device in question), it gives out that they cannot be bandlimited, according to the time-frequency relationship. Thus, aliasing would happen and lead to sampling distortions.

2) Integration effect or aperture effect. This effect results from the fact that the sample is obtained as time average within a sampling region, rather than just being equal to the signal value at the sampling instant. The integration effect is readily noticeable in photography when the exposure is too long and creates a blur in the image. An ideal camera would have an exposure time of zero. In a capacitor-based sample and hold circuit, the integration effect is introduced because the capacitor cannot instantly change voltage thus requiring the sample to have non-zero width.

3) Jitter or deviation from accurate sampling time intervals due to limitation of the physical device.

4) Noise, including thermal sensor noise, analog circuit noise, etc.

5) Slew rate limit error. It can be caused by inability when the device's output changes sufficiently fast.

6) Quantization as a consequence of the finite precision of words that represent the converted values. In practice, the measurement matrix can only be able to incorporate finite frequencies. The estimated spectrum can only be represented with respect to the finite discrete frequencies. However, the primary signal's spectrum can be in any position within the sampling frequency range. When the active primary signals are not located in the divided frequencies, quantization error would occur and lead to sampling distortion. It is more serious when the sampling frequency band is wide.

7) Error due to other non-linear effects when converting input voltage to output value can also result in sampling error.

All the above effects can give birth to sampling distortion in AIC, but the traditional sparse spectrum model (5) does not consider sampling distortion. Performance degrade would be caused by the inaccurate signal model. To make the sparse spectrum model more practical, the sampling distortion can be modeled as a bounded additive noise on the measurement matrix. The new sparse spectrum model with sampling distortion can be formulated as:

$$\mathbf{y} = \mathbf{\Psi}\mathbf{y}_t = \mathbf{\Psi}\mathbf{\Phi}\mathbf{b} = \mathbf{A}\mathbf{b}, \quad \mathbf{b} \in \mathbf{R}^N, \tag{6}$$

$$\mathbf{B} = \mathbf{A} + \mathbf{V}, \tag{7}$$

Where $M \times 1$ vector $\mathbf{y}$ is the random samples, the real $M \times N$ matrix $\mathbf{A}$ is unknown, $\mathbf{B}$ is the $M \times N$ observed matrix with an additive noise $\mathbf{V}$, and the $N \times 1$ vector $\mathbf{b}$ is sparse with most of its elements equal zero or close to zero.

Without loss of generality, we assume that $\mathbf{V}$ is deterministic and satisfies:

$$\|\mathbf{V}\|_\infty \leq \delta, \tag{8}$$

where $\delta > 0$, and $\|\mathbf{V}\|_\infty$ stands for the $\ell_\infty$ norm for the matrix $\mathbf{V}$, which gives the maximum of $\sum_{i=1}^{N}|v_{i,j}|$ for all $j$. The $v_{i,j}$ is the element of matrix $\mathbf{V}$ in $i$-th column and $j$-th row. The parameter $\delta$ gives a restriction on the bound of the observed measurement matrix noise $\mathbf{V}$.

If $\mathbf{V}$ is random, condition (8) can also be guaranteed with a probability close to 1; and the values of $\delta$ can be indicated correspondingly. Thus, the result that we prove for deterministic $\mathbf{V}$ is extended in a trivial way to their random version satisfying this assumption.

## IV. Basis Pursuit Based Compressive Wideband Spectrum Sensing

CS theory asserts that, if a signal has sparse representation in a certain space, one can use the random measurement to obtain the samples and reconstruct it with overwhelming probability by optimization techniques, as stated in section II. The required random samples for recovery are far fewer than Nyquist sampling requires.

To find the unoccupied spectrum for the secondary access, the signal in the monitoring band is down-converted to baseband and sampling the resulting analogue signal through an AIC that produces samples at a rate below the Nyquist rate.

Now we estimate the frequency response of *y(t)* from the measurement *y* based on the transformation equality $\boldsymbol{y} = \boldsymbol{S}_c \boldsymbol{F}_N^{-1} \mathbf{r}$, where $\mathbf{r}$ is the $N \times 1$ frequency response vector of signal *y(t)*, $\boldsymbol{F}_N$ is the $N \times N$ Fourier transform matrix, and $\boldsymbol{S}_c$ is the $M \times N$ matrix which is obtained by randomizing the column indices and getting the first $M$ columns. The real measurement matrix can be formulated as:

$$\mathbf{A} = \mathbf{S}_c \mathbf{F}_N^{-1}, \tag{9}$$

In practice, as it states in section III, the sampling may suffer various physical effects and get the measurement matrix distortion as (7), and the obtained observed measurement matrix should be

$$\mathbf{B} = \mathbf{S}_c \mathbf{F}_N^{-1} + \mathbf{V}, \tag{10}$$

With the observed measurement matrix **B** and the random measurement vector **y**, the CWSS can be formulated as:

$$\hat{\mathbf{r}} = \arg\min_{\mathbf{r}} \|\mathbf{r}\|_0, \tag{11}$$

$$\text{s.t. } \mathbf{y} = \mathbf{Br}, \tag{12}$$

where $\|\mathbf{r}\|_0$ is the $\ell_0$ norm which counts the number of nonzero entries of the vector **r**. The minimization of the $\ell_0$ norm is the optimal to describe the sparse distribution. However, it is not a convex function and the formulated CWSS (11)(12) is not a convex programming.

To convert the CWSS model to a convex programming, the $\ell_0$ norm is relaxing to the $\ell_1$ norm, and the Basis pursuit (BP) is used to performance CWSS. The globe optimal sparse solution can be obtained. The BP based CWSS (BP-CWSS) can be formulated as [9][15]:

$$\hat{\mathbf{r}} = \arg\min_{\mathbf{r}} \|\mathbf{r}\|_1, \tag{13}$$

$$\text{s.t. } \mathbf{y} = \mathbf{Br}, \tag{14}$$

where $\|\mathbf{r}\|_1 = \sum_i |r_i|$ is the $\ell_1$ norm of the vector $\mathbf{r} = [r_1, r_2, \ldots, r_N]^T$. This problem is a second order cone program (SOCP) and thereby can be solved efficiently using standard software packages. It finds the smallest L1 norm of coefficients among all the decompositions that the signal is decomposed into a superposition of dictionary elements. It is a decomposition principle based on a true global optimization.

The standard BP does not allow sampling noise. In contrast, LASSO, also called basis pursuit denoising (BPDN), has additional advantage of noise suppression [25]. It is a shrinkage and selection method for linear regression. It minimizes the usual sum of squared errors, with a bound on the sum of the absolute values of the coefficients. To get more accuracy, we can reformulate the CWSS based on LASSO as

$$\hat{\mathbf{r}} = \arg\min_{\mathbf{r}} \|\mathbf{r}\|_1, \quad (14)$$

$$\text{s.t.} \ \|\mathbf{Br} - \mathbf{y}\|_2 \leq \mu, \quad (15)$$

where $\mu$ is a parameter bounding the amount of noise in the data. As it directly uses the sparsity to recover the wide frequency band spectrum by LASSO with reduced measurements, we named it as LASSO based compressive wideband spectrum sensing (LASSO-CWSS). The problem of LASSO is a quadratic programming problem or more generally, a convex optimization problem, which can be tackled by standard numerical analysis algorithms. The solution for it has been well investigated [25]-[28]. A number of convex optimization software, such as cvx [29], SeDuMi [30] and Yalmip [31], can be used to solve the problem.

## V. ROBUST COMPRESSIVE WIDEBAND SPECTRUM SENSING WITH SAMPLING DISTORTION

Traditional ways for CWSS are designed for the sparse signal model (5). It formulates with respect to the real measurement matrix without any sampling distortion. However, in practice the continuous signal is sampled using the AIC, which is a non-ideal device with various physical effects leading to the deviations from the theoretically perfect reconstruction capabilities. Conditions (14) and (16) would mismatch the more practical sparse spectrum signal model (6) (7). The sampling distortion would lead to the performance degeneration of the BP-CWSS and the LASSO-CWSS.

In cognitive radio, the spectrum sensing accuracy is vitally important. The inaccurate wideband spectrum sensing would lead to missed detection or false detection. The missed detection can make the cognitive radio occupy the spectrum which the active primary users are using. Thus, both the cognitive radio and the primary user can cause and get interference from each other. Both of their communication quality can seriously degrade. On the other hand, false detection would exclude the cognitive radio to take the spectrum where no primary user is using. As both the cognitive radio and primary user are not taking the spectrum, the basic ideal that using cognitive radio to access the unused spectrum is not realized. Spectrum waste would still exist.

To mitigate the performance degeneration caused by the sampling distortion, we incorporate the sampling distortion information in the sparse wideband spectrum recovery algorithm. Taking advantage of the sparse spectrum signal model (6) (7), we can make up the performance degeneration caused by the sampling distortion in sparse spectrum reconstruction algorithm. To get the robust sparse wideband spectrum estimation algorithm, we can reformulate the square estimation error as:

$$\begin{aligned} \|\mathbf{y} - \mathbf{B}\boldsymbol{\theta}\|_2 &= \|\mathbf{y} - (\mathbf{A} + \mathbf{V})\mathbf{r}\|_2 \\ &= \|\mathbf{A}\mathbf{r} - (\mathbf{A} + \mathbf{V})\mathbf{r}\|_2 \\ &= \|\mathbf{V}\mathbf{r}\|_2 \end{aligned} \qquad (17)$$

Here we define the row vector $\mathbf{v}_i$, i = 1, 2, ... , M of the matrix $\mathbf{V}$ as:

$$\mathbf{V} = \begin{bmatrix} \mathbf{v}_1 \\ \mathbf{v}_2 \\ \vdots \\ \mathbf{v}_M \end{bmatrix}. \qquad (18)$$

Then (17) can be reformulated as:

$$\|\mathbf{V}\mathbf{r}\|_2 = \sqrt{\sum_{m=1}^{M} |\mathbf{v}_m \mathbf{r}|^2}, \qquad (19)$$

where $|\bullet|$ means the modulus operator of a scalar. It is easy to prove that

$$|\mathbf{v}_m \mathbf{r}| = |v_{m,1} r_1| + |v_{m,2} r_2| + \cdots + |v_{m,N} r_N|$$
$$\leq \left(|v_{m,1}| + |v_{m,2}| + \cdots + |v_{m,N}|\right)\left(|r_1| + |r_2| + \cdots + |r_N|\right), \quad (20)$$
$$= \|\mathbf{v}_m\|_1 \|\mathbf{r}\|_1$$

where $v_{m,i}$, i = 1, 2, … , N, is the i-*th* element of the vector $\mathbf{v}_m$; and $r_i$, i = 1, 2, … , N, is the i-*th* element of the vector $\mathbf{r}$. Taking (20) into (19), we can get

$$\|\mathbf{V}\mathbf{r}\|_2 \leq \sqrt{\sum_{m=1}^{M} \left(\|\mathbf{v}_m\|_1 \|\mathbf{r}\|_1\right)^2}. \quad (21)$$

Then, obviously we have

$$\|\mathbf{V}\mathbf{r}\|_2 \leq \sqrt{\sum_{m=1}^{M} \left(\|\mathbf{v}\|_{1,\max} \|\mathbf{r}\|_1\right)^2}$$
$$= \sqrt{M \left(\|\mathbf{v}\|_{1,\max} \|\mathbf{r}\|_1\right)^2}, \quad (22)$$
$$= \sqrt{M} \|\mathbf{V}\|_\infty \|\mathbf{r}\|_1$$

where $\|\mathbf{v}\|_{1,\max}$ means the maximum one of all the values of the $\|\mathbf{v}_i\|_1$, i = 1, 2, … , M. With the condition (8), we can get:

$$\|\mathbf{V}\mathbf{r}\|_2 \leq \sqrt{M} \delta \|\mathbf{r}\|_1. \quad (23)$$

Combining (17) and (23), we get the anti-sampling-distortion constraint (ASDC), it can be formulated as:

$$\|\mathbf{y} - \mathbf{B}\mathbf{r}\|_2 \leq \sqrt{M} \delta \|\mathbf{r}\|_1. \quad (24)$$

It is a relaxing version of the standard square error bound constraint and gives the relationship between the square error and sparsity measure of the estimated signal with the sampling distortion parameter incorporated.

Combining the ASDC with the sparse constraint, we get the anti-sampling-distortion compressive wideband spectrum sensing (ASD-CWSS) as:

$$\begin{aligned} & \min \|\mathbf{r}\|_1 \\ & \text{s.t. } \|\mathbf{y} - \mathbf{B}\mathbf{r}\|_2 \leq \sqrt{M} \delta \|\mathbf{r}\|_1 \end{aligned}. \quad (25)$$

It can be reformulated as

$$\min(t)$$
$$\text{s.t. } \|\mathbf{y} - \mathbf{Br}\|_2 \leq \sqrt{M}\delta t. \qquad (26)$$
$$\|\mathbf{r}\|_1 < t$$

Obviously, (26) is a convex programming and can be solved by a host of numerical methods in polynomial time. Similar to the solution of LASSO-CWSS, the optimal **r** of ASD-CWSS can also be obtained efficiently using some convex programming software packages. Such as cvx [29] and SeDuMi [30], Yalmip [31], etc.

## VI. SIMULATION RESULTS

Numerical experiments are presented to illustrate performance of the proposed ASDC-CWSS for cognitive radio. Here we consider a base band signal with its frequency range from 0Hz to 500MHz. Among the whole monitoring band, the noise level is set to be in random Gaussian distribution between 0 and 10, Four primary signals are located at 30MHz – 70MHz, 120MHz – 180MHz, 300MHz – 340MHz, 420MHz – 460MHz. And their corresponding normalized Power Spectrum Density (PSD) levels fluctuate in the range of 100-140, 70 - 110, 130 - 170, and 110 - 150. The primary signals with random phase are contaminated by a zero-mean additive Gaussian white noise (AGWN) which makes the signal noise ratio (SNR) be 13dB. Then the wideband spectrum in normalized as Fig. 2 shows. The observed measurement matrix noise **V** is a complex Gaussian random matrix with the bound of the modulus of elements being $\delta = 0.7$.

Here we take the contaminated signal as the received signal y(t). As CS theory suggests, we sample y(t) randomly at half the amount of Nyquist sampling. The resulted sub-sample vector by the proposed structure as Fig. 1 demonstrated is denoted as *y*. To make contrast, with the same number of samples, power spectrums with different methods are given out in Fig. 3 and

Fig. 4. Fig. 3 shows the spectrum estimation result by LASSO-CWSS where $\mu$ is chosen to be $0.1\|y\|_2$ with 1000 trials averaged. Fig. 4 shows the result of the proposed ASD-CWSS also with 1000 trials averaged. Fig. 3 and Fig. 4 show that the ASD-CWSS gives a better reconstruction performance with the same amount of samples. In detail, it is shown that there are too many fake spectrum points in the subbands in the absence of active primary signals in Fig. 3 for the LASSO-CWSS. The noise levels from LASSO are quite high in the whole monitoring band. However, for the ASD-CWSS, the four occupied bands clearly show up. The noise levels in the inactive bands are quite low; and the variation of the PSD levels in the boundaries of estimated spectrum are quite abrupt and correctly in accordance with the generated sparse spectrum, which would enhance the wavelet edge detection performance much [32]. Therefore, the proposed ASD-CWSS outperforms the LASSO-CWSS with a much more accurate spectrum reconstruction as shown in Fig. 3 and Fig. 4.

Apart from the edge detection, energy detection is the most popular spectrum sensing approach for cognitive radio. To test compressive wideband spectrum sensing by energy detection, 1000 Monte Carlo simulations are conducted with same parameters given in the last simulation to show the results of average energy in each section of the divided spectrum vector. Here the simulated monitoring band is divided into 9 sections as Fig 2. The total energy with each CWSS method is normalized. Table 1 presents the average energy in each subband with the corresponding CWSS methods. For the ASD-CWSS, it is obvious that the estimated noise energy of inactive bands is much smaller that the LASSO-CWSS. To quantify the ASD-CWSS's performance gain against LASSO-CWSS, after normalizing the total energy with each method, we define the energy enhancement ratio (EER) for the $k$-th subband as:

$$EER(k) = \begin{cases} \dfrac{E_k^N - E_k^S}{E_k^S}, & \text{for active subbands} \\ \dfrac{E_k^S - E_k^N}{E_k^S}, & \text{for inactive subbands} \end{cases}, \qquad (27)$$

where $E_k^N$ represents the energy in the *k*-th subband range obtained by the newly proposed ASD-CWSS, and $E_k^S$ represents the energy vector in the *k*-th subband range obtained by the LASSO-CWSS.

This performance function can quantify how much energy increased to enhance the probability of correct energy detection of the active primary bands and how much denoising performance is enhanced. The EER in Table 1 clearly tells the how great the improvement achieved by the proposed ASD-CWSS method. Although there is a subband with declined energy, the other active subbands have a large energy enhancement and the denoising performance in inactive subbands is improved much.

By the way, in 1000 Monte Carlo simulations, on our hardware condition, the average computing periods for the LASSO-CWSS and the ASD-CWSS are 71.7342 seconds, 82.3906 seconds respectively. The latter way has a little longer computing periods than that of the former one, but the ASD-CWSS gets considerable enhancement of spectrum estimation accuracy with the same number of samples, which qualifies it as an more suitable candidate for wideband spectrum sensing.

## VII. Conclusion

Spectrum scarsity is a big problem for mobile communication. Cognitive radio is the most promising way to deal with it. Wideband spectrum sensing is an essential functionality for cognitive radio to detect the secondary spectrum access opportunity without any harmful interference to the primary user. Too high sampling requirement is the most difficult problem for the robust and fast wideband spectrum sensing. Exploiting the inherent sparsity in the wide monitoring band for cognitive radio, CS based sparse signal estimation can perform wideband spectrum sensing with sub-Nyquist random samples which is obtained by AIC. However, traditional sparse signal recovery algorithms do not consider sampling distortion in AIC. Here we model the sampling error to be a bounded additive noise. To get the robust CWSS against sampling distortion, the standard square error bound constraint is replaced by an ASDC in sparse signal recovery algorithm LASSO. The proposed ASDC is in terms of $\ell_1$ and $\ell_2$ norms of the estimated spectrum and the error bound in sampling distortion. It is a relaxing version of the standard square error bound constraint and gives the relationship between the square error and sparsity of the estimated signal with the sampling distortion parameter incorporated. Numerical experiments show that the proposed ASD-CWSS has performance superior to the LASSO-CWSS in accuracy and denoising capacity. The proposed ASD-CWSS way can be a more suitable candidate for wideband spectrum sensing.

In the future, compressive wideband spectrum sensing can be introduced to the cognitive radio for muti-antenna system and wireless sesnor networks[33][34]. Besides, some more apriori information of spectrum distribution can be incorparted to enhance the spectrum sensing accurnacy. Greedy algorithm can also be used to solve the ASD-CWSS problem to get a faster spectrum sensing.

ACKNOWLEDGMENT

The authors would like to thank Dr. Ying Zhang who performs carefully words and grammar checking for the whole paper. This work was supported in part by the National Natural Science Foundation of China under grant 60772146, the National High Technology Research and Development Program of China (863 Program) under grant 2008AA12Z306 and in part by Science Foundation of Ministry of Education of China under grant 109139.

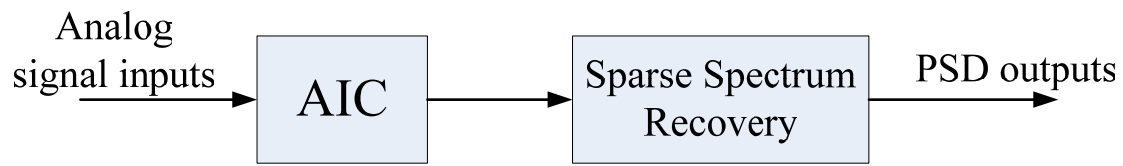

Fig. 1 The proposed compressive wideband spectrum sensing structure

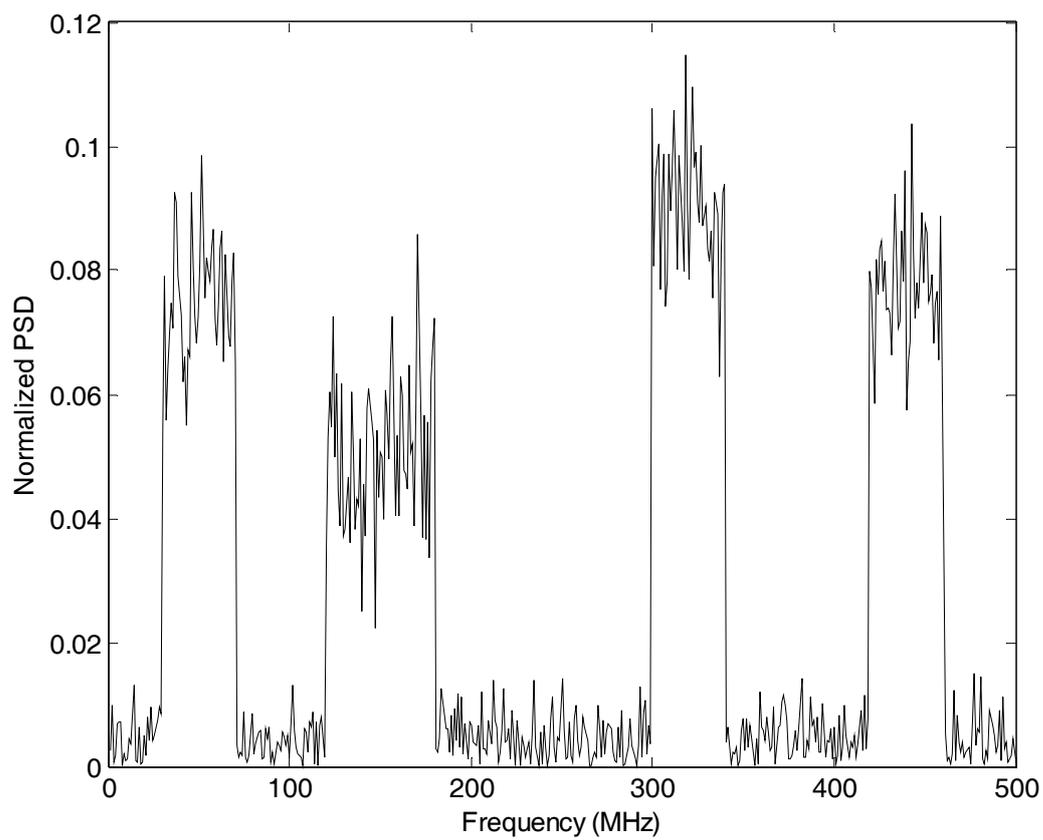

Fig 2. The spectrum of noiseless active primary signals in the monitoring band.

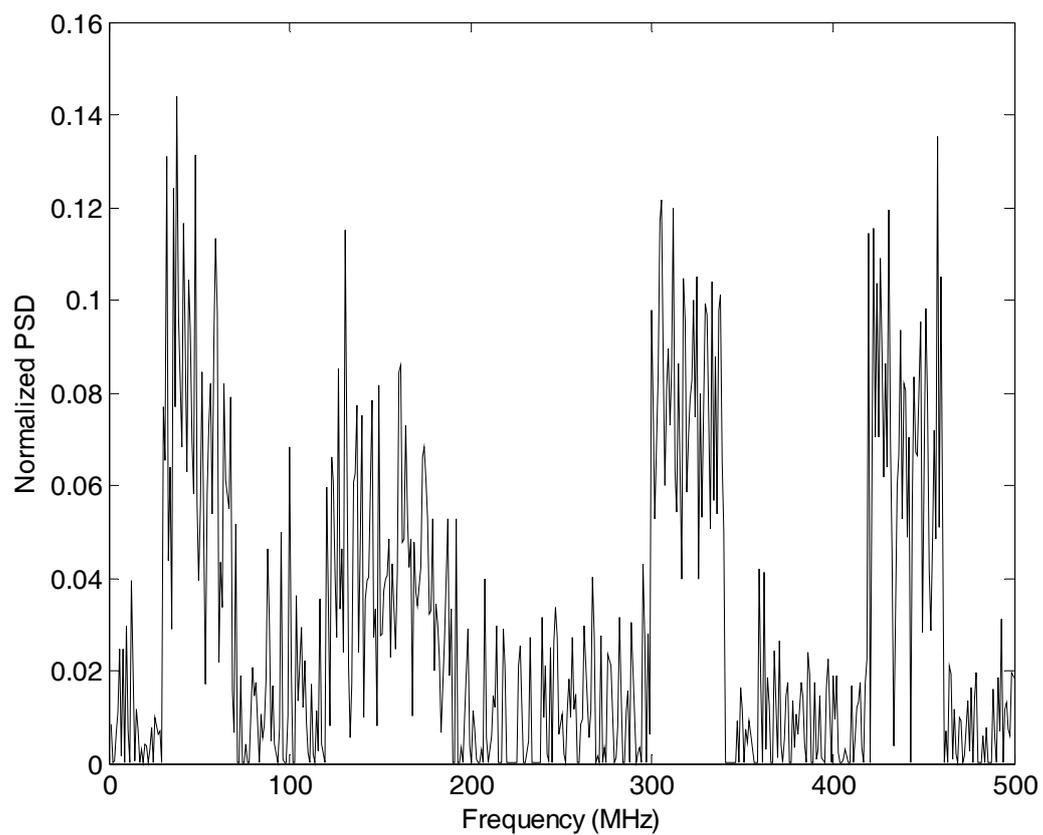

Fig. 3 The compressive wideband spectrum estimation by the LASSO-CWSS.

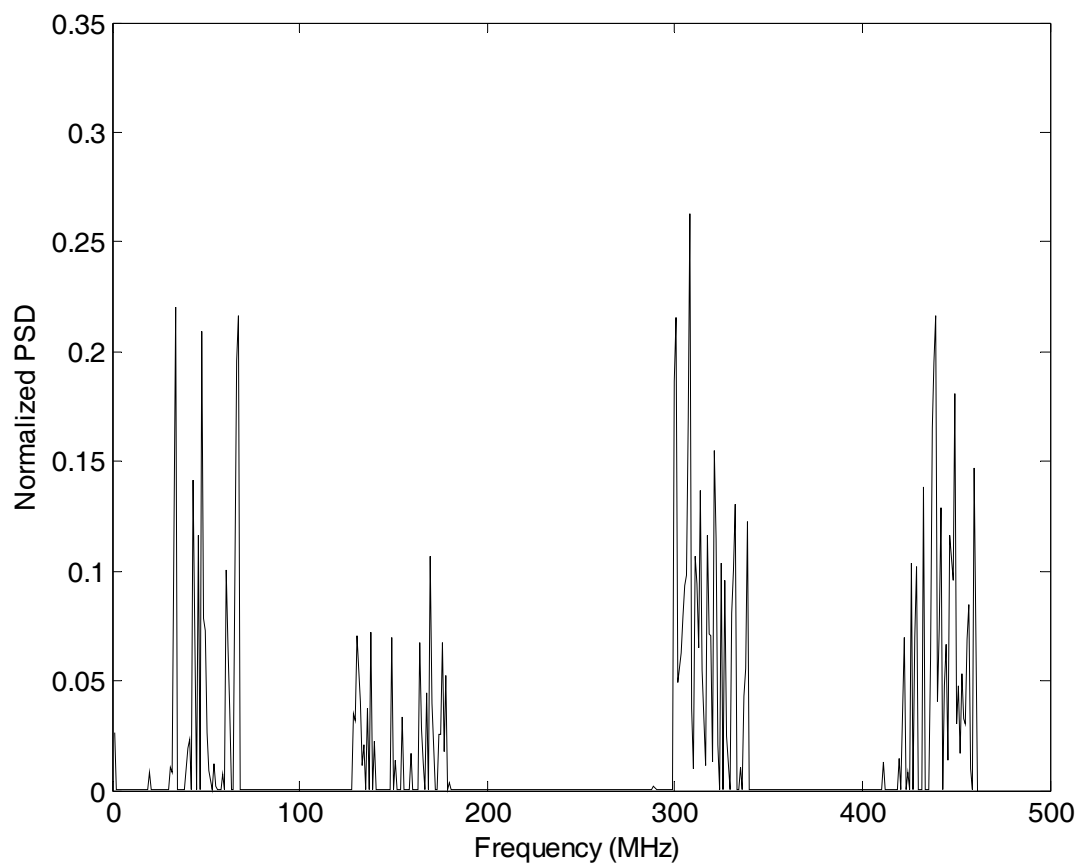

Fig. 4 The compressive wideband spectrum estimation by the proposed ASD-CWSS.

|            | 1      | 2      | 3      | 4      | 5      | 6      | 7      | 8      | 9      |
|------------|--------|--------|--------|--------|--------|--------|--------|--------|--------|
| LASSO-CWSS | 0.0298 | 0.1601 | 0.0335 | 0.1854 | 0.0985 | 0.2052 | 0.0685 | 0.1836 | 0.0352 |
| ASD-CWSS   | 0.0036 | 0.2918 | 0.0000 | 0.0990 | 0.0002 | 0.3552 | 0.0013 | 0.2484 | 0.0006 |
| EER        | 0.8792 | 0.8222 | 1.0000 | 0.4662 | 0.9979 | 0.7307 | 0.9811 | 0.3524 | 0.9830 |

Table 1  The total energy in each subband with two compressive wideband spectrum sensing methods and the EBR.